\documentclass[10pt,aps,prl,groupedaddress,twocolumn, nofootinbib]{revtex4-1}

\usepackage[utf8]{inputenc}
\usepackage{hyperref}
\usepackage{graphicx}  
\usepackage{dcolumn}   
\usepackage{bm}        
\usepackage{amssymb,amsmath}   
\usepackage{uri}
\usepackage[dvipsnames]{xcolor}

\hypersetup{
    bookmarks=true,         
    unicode=false,          
    pdftoolbar=true,        
    pdfmenubar=true,        
    pdffitwindow=false,     
    pdfstartview={FitH},    
    pdfauthor={Kevin Multani},     
    colorlinks=true,       
    linkcolor=blue,          
    citecolor=red        
}

\begin{document}

\title{Stochastic resetting antiviral therapies prevents drug resistance development}
\author{Angelo Marco Ramoso$^{1}$, Juan Antonio Magalang$^{1}$,  Daniel~S\'anchez-Taltavull$^{2}$, Jose Perico Esguerra$^1$, and \'Edgar Rold\'an$^3$}
\affiliation{$^1$ Theoretical Physics Group, National Institute of Physics, University of the Philippines, Diliman, Quezon City 1101, Philippines\\
$^2$ Department for Visceral Surgery and Medicine, Bern University Hospital, University of Bern, Switzerland\\
$^3$ICTP - The Abdus Salam International Centre for Theoretical Physics, Strada Costiera 11, 34151, Trieste, Italy}
\date{\today}


\begin{abstract}
We study minimal mean-field  models of viral drug resistance development in which the efficacy of  a  therapy is  described by a one-dimensional stochastic resetting process with mixed reflecting-absorbing boundary conditions. We derive analytical expressions for the mean survival time for the virus to develop complete resistance to the drug. 
We show that the optimal therapy resetting rates that achieve a minimum and maximum mean  survival times undergo a second and first-order phase transition-like behaviour as a function of the therapy efficacy drift. We illustrate our results with  simulations of a population-dynamics model of HIV-1 infection.
\end{abstract}

\maketitle

 Antiviral and antiretroviral therapies are continuously being developed to tackle viral diseases~\cite{lichterfeld2011treating}. Because of  viral evolution, a therapy that is effective today may not remain effective forever~\cite{clutter2016hiv}. This process is known as drug resistance development and it is of special importance in chronic infections such as HIV-1 \cite{pillay1998antiviral,perelsonnelsonhiv-1,sanchez-taltavull2016hiv-1eradication}, and also for herpes and hepatitis \cite{strasfeld2010antiviral}.
Viral evolution results into an increase in the number of infected cells, leading to a failure of the normal functions of the infected patient that can result in their  death~\cite{clutter2016hiv}.
 Because viruses replicate and mutate rapidly, and mutation is a random process, viral evolution is intrinsically noisy~\cite{fabreti2019stochastic,manrubiaviralevolution}. 
 When possible, practitioners overcome this situation by changing the therapy given to  the patient.
  Changes in antiviral therapies can occur because of a detected viral resistance~\cite{pillay1998antiviral,wensing2019update}, or for purely stochastic reasons e.g. the appearance in the market of a therapy that is cheaper or has less secondary effects.
  It remains an open yet challenging problem to characterize and design  therapy change protocols that ensure large patient survival times that 
  are robust to   drug-resistance  fluctuations.


Stochastic models have been used to study viral evolution~\cite{fabreti2019stochastic,wensing2019update,manrubiaviralevolution,tria2005minimal,nelson2006stochastic}. Examples include the usage of branching process to describe viral persistence and extinction~\cite{fabreti2019stochastic,manrubiaviralevolution}, 
and   population dynamics models    to assess 
the impact 
of a treatment in drug resistance development in the context of HIV-1~\cite{zitzmann2018mathematical}.
A model that described RNA virus evolution  as a diffusion process in a fitness landscape~\cite{tsimring1996rna} was able to explain the experimentally-observed    rapid initial growth  
followed by a slower stage of linear growth in the logarithm of fitness  of  clone colonies of   vesicular stomatitis virus~\cite{novella1995exponential, holland1991quantitation}. 
In the same vein, the notion of a fluctuating therapy efficacy can be used  to  describe  the evolution of therapy protocols. 



 \begin{figure}[t]
\includegraphics[width=1.0\linewidth]{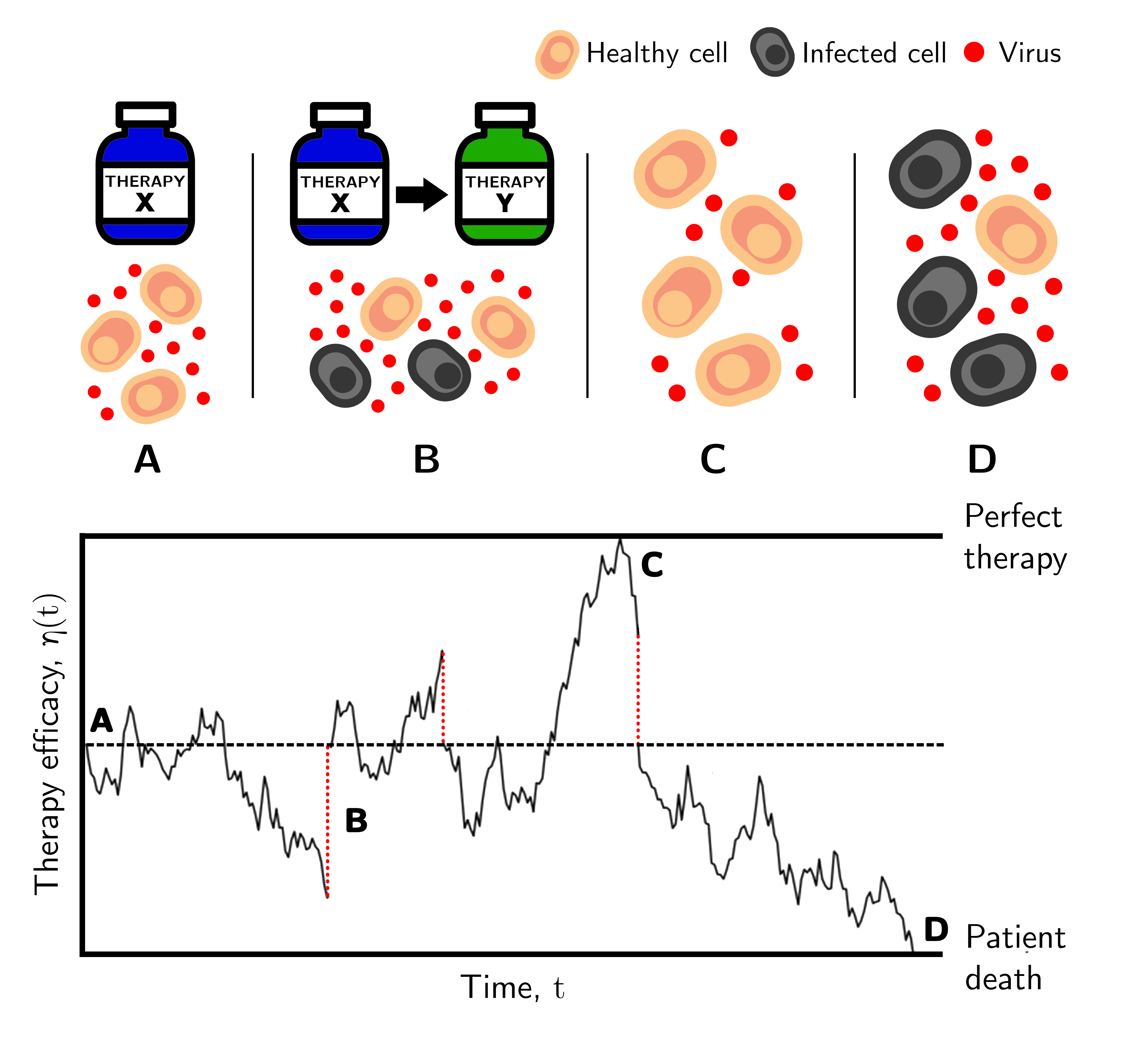}
\caption{Illustration of the stochastic-resetting model of antiviral therapy efficacy. Top: Sketch of the main ingredients of the model. Bottom: Sample trajectory of the therapy efficacy as a function of time, with A-D illustrating key events during a single realization of the process. A) Initial condition, for a $50\%$ efficient therapy $\eta_0=1/2$. B) Change of the current therapy at a random time (stochastic resetting) by a new one with $50\%$ efficacy. C) Reflecting boundary condition modelling the maximum possible therapy efficacy $\eta=1$. D) Absorbing boundary condition yielding to the dead of the patient due to a completely inefficient therapy  $\eta=0$.  }
\label{fig:1}
\end{figure}

Viral evolution can be modelled as a stochastic diffusion process 
 involving incremental changes in efficacy ---due to viral mutation--- punctuated by sudden changes in efficacy ---due to changes in therapy---. Diffusions with stochastic resetting   \cite{evans2011diffusion,evans2020topicalreviewstochasticresetting,bhat2016stochastic, kusmierz2015optimal,pal2019firstpassageresettinginterval,pal2019landauphasetransitionsstochasticresetting,durang2019firstpassagestochasticresettingboundeddomain,christou2015diffusionresettingbounded,pal2017firstpassagerestart,chechkin2018searchresettingunifiedrenewal,ray2019pecletdriftdiffusion,basu2019symmetric} thus  provide a promising framework to model therapy evolution. This framework has been instrumental to describe biophysical processes with sudden changes, namely RNA polymerase backtracking~\cite{roldan2016stochastic,tucci2020controlling}, receptor dynamics~\cite{mora2015physical},  cell crawling~\cite{bressloff2020stochastic}, and population dynamics~\cite{mercado2018lotka,da2018interplay,garcia2019linking}, 
   see~\cite{evans2020topicalreviewstochasticresetting} for a review of applications.  It remains unclear  how in multiscale processes   stochastic resetting of a slow variable (e.g. a therapy efficacy)
    affects a fast variable describing the dynamics of a time-varying population (e.g. cells and viruses). 

 In this Letter, we introduce a  stochastic-resetting model for the evolution of the efficacy of an antiviral therapy, and study its evolution under different treatment protocols.  We  first describe the therapy efficacy as a one-dimensional resetting biased diffusion model with mixed absorbing and reflecting boundaries, calculate an exact analytical expression for the mean first passage time, and test our result  by comparing it with Langevin dynamics simulations.  We  then discuss 
 the clinical effects of our viral evolution model by coupling the stochastic resetting biased-diffusion therapy efficacy  to a population dynamics model of HIV-1 chronic disease.

 

We describe the efficacy of an antiviral treatment as a bounded stochastic process $0\leq  \eta(t)\leq 1$ where $\eta=0$ and $\eta=1$ correspond  to a completely ineffective and  completely effective  therapy, respectively. We assume that treatments that stop working lead to the death of the patient, i.e.  $\eta=0$ is an absorbing boundary. On the other hand $\eta=1$ is a reflecting boundary set at the maximum $100\%$  therapy efficacy. 
 We model the evolution of therapy efficacy $\eta(t)$ as a  biased  random walk or drift-diffusion  process in "efficacy space" with diffusion coefficient $D$  and drift $v$. 
  The therapy is often biased in the direction of lesser efficacy, i.e. the  drift is negative $v<0$, due to the fact that  viruses  develop  resistance to the existing therapies that survive in the long run. Furthermore, we  complement the biased diffusion model with a resetting protocol that switches  the therapy efficacy to a value $\eta_0$ instantaneously at random Poissonian times with rate $r$. 
 Such therapy "resetting" events can be due to the  introduction of a new dose of drug, the discovery of more effective variants of the therapy, etc. See Fig.~\ref{fig:1} for an illustration of the model and a sample stochastic trajectory of the therapy efficacy.

The  evolution of the model can be described by the Fokker-Planck equation with source terms
\begin{equation}
    \partial_t P + \partial_\eta (vP - D\partial_\eta P) = -r P+ r \delta (\eta-\eta_0),
    \label{eqn:diffeqn}
\end{equation}
where $P \equiv P (\eta, t|\eta_0,0)$ is the conditional probability  density that the therapy is at $\eta$ at time $t$ given that its initial value (at which it is reset at rate $r$) was $\eta_0$. 
The dynamics is complemented by a absorbing boundary condition at $\eta=0$, $\lim_{\eta \rightarrow 0^+} P (\eta, t) = 0$ and a zero-flux reflecting boundary condition at $\eta=1$, $J (1, t) = 0$ where
 $J (\eta, t)= v P (\eta, t) - D \partial_\eta  P (\eta, t) $ is the  probability current.

 


 In the following, we derive analytical expressions for the finite-time survival probability and the mean survival time. We denote by {\em survival time} the first-passage time of the therapy efficacy to reach the absorbing boundary $\eta=0$. To this aim, we first make use of a  relation between the finite-time survival probability with resetting $S_r(\eta_0,t)$ and the survival probability without resetting $S_0(\eta_0,t)$~\cite{pal2019firstpassageresettinginterval} 
\begin{equation}
S_r (\eta_0, t)=e^{- r t}S_0 (\eta_0, t)+r \int_0^t\! \text{ d} \tau e^{- r \tau} S_0 (\eta_0, \tau) S_r (\eta_0, t - \tau).
\label{eq:2}
\end{equation}
From Eq.~\eqref{eq:2} we show that the Laplace transform of the first passage probability with resetting $\Tilde{S}_r (\eta_0,s)$  and without resetting $\Tilde{S}_0 (\eta_0,s)$ are related through the identity (see Supplemental Material)
\begin{equation}
\Tilde{S}_r (\eta_0,s) = \dfrac{( r + s ) \Tilde{S}_0 (\eta_0,s + r)}{s + r \Tilde{S}_0 (\eta_0,s + r)}.
\end{equation}
We show that the Laplace transform of the first passage survival probability without resetting is given by
\begin{equation}
    \Tilde{S}_0(\eta_0,s)= e^{-\frac{\eta_0 v}{2D}}\frac{2D\omega\cosh[\omega(\eta_0-1)] + v\sinh[\omega(\eta_0-1)]}{2D\omega\cosh\omega - v\sinh \omega}
    \label{eqn:fs}
\end{equation}
where $\omega=\sqrt{v^2+4Ds}/2D$. 
Substituting Eq.~\eqref{eqn:fs} in Eq.~\eqref{eq:2} and using the relation $\langle \tau_r \rangle=-\lim_{s\to 0} \partial\Tilde{S}_r(\eta_0,s)/\partial s$ we obtain the following analytical expression for the mean first-passage time
\begin{equation}
\langle \tau_r \rangle    =\frac{\phi(\text{Pe},\Omega,\eta_0)}{r \left\{ \text{Pe} \sinh \Big[ \Omega \big( \eta_0 - 1 \big) \Big]  + \Omega \cosh \Big[ \Omega \big( \eta_0 - 1 \big) \Big]\right\}^2},
\label{eq:mfpt}
\end{equation}
where the non-trivial function  $\phi(\text{Pe},\Omega,\eta_0)=\text{Pe}^2 \{ e^{\text{Pe}\eta_0} \cosh [ \Omega ( \eta_0 - 2 ) ]- \cosh [ 2 \Omega ( \eta_0 - 1 )] \}+\text{Pe} \Omega\, \{ e^{\text{Pe}\eta_0} \sinh [ \Omega \big( \eta_0 - 2 \big) ]- \sinh [ 2 \Omega ( \eta_0 - 1 )] \}+  e^{\text{Pe}\,\eta_0} \cosh [ \Omega \eta_0 ] - \cosh [ 2 \Omega ( \eta_0 - 1 ) ]+ e^{\text{Pe}\,\eta_0} \cosh [ \Omega ( \eta_0 - 2 )] -1 $ depends on the model parameters through the dimensionless quantities  $
\text{Pe} = v/2 D, \quad \Omega = \sqrt{v^2 + 4 D r}/2 D$, and $\xi = r/2D$.

\begin{figure}[t]
\includegraphics[width=0.9\linewidth]{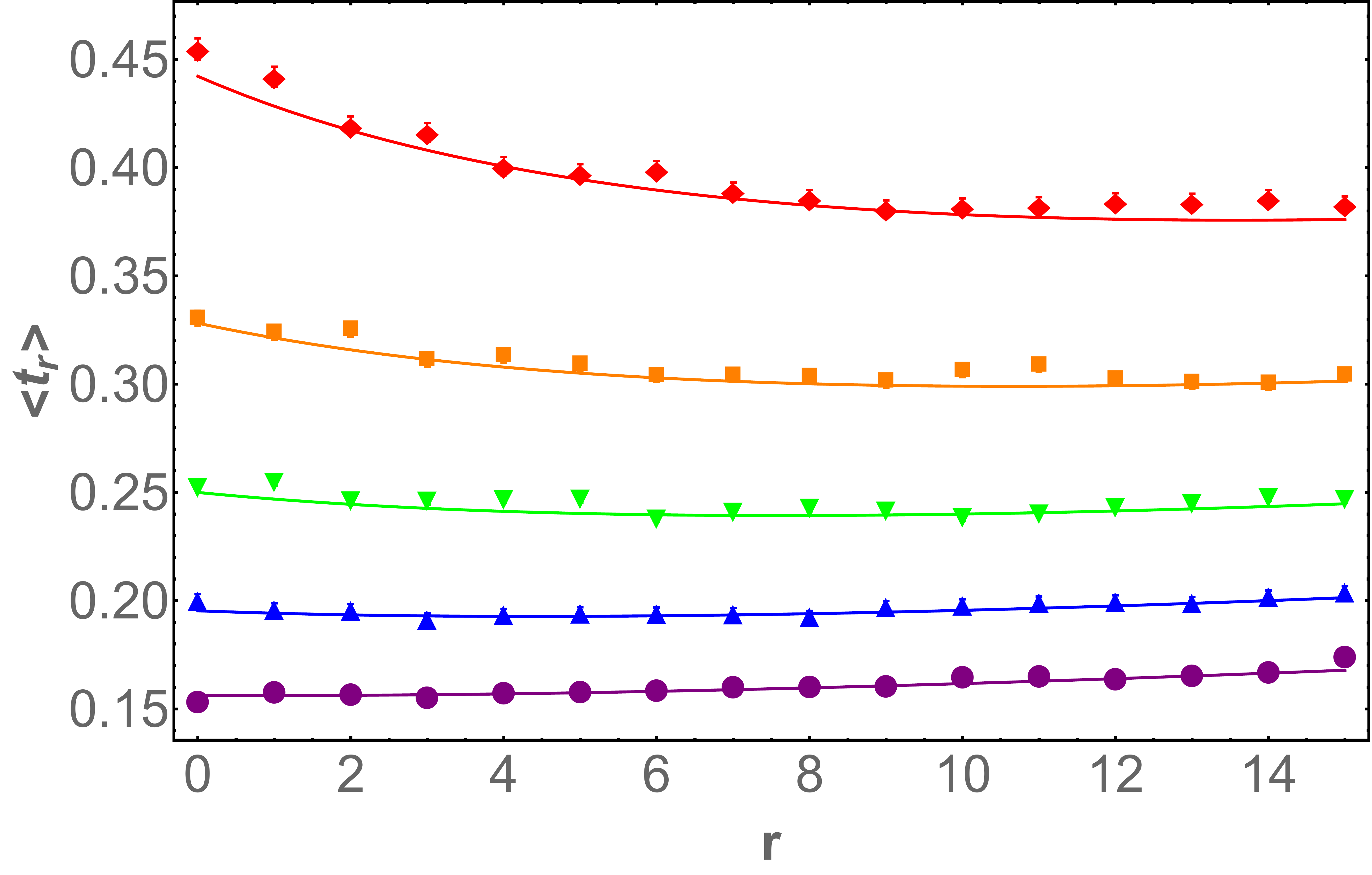}
\caption{Mean survival time for a therapy to become completely inefficient, as a function of the therapy resetting rate: numerical simulations (symbols) and analytical  results (lines) given by Eq.~\eqref{eq:mfpt}. The data corresponds to parameter values diffusion $D = 1.5$, initial efficacy $\eta_0=1/2$ and different values of drift $v$: $v=2$~({\color{red}{$\blacklozenge$}}), $v=1$ ({\color{orange}{$\blacksquare$}}) $v=0$ ({\color{green}{$\blacktriangledown$}}), $v=-1$ ({\color{blue}{$\blacktriangle$})}, and $v=-2$ ({\large \color{violet}{$\bullet$}}). For all parameter values, the simulations were done using  Euler's numerical integration with  parameters: $10^5$ number of simulations, each with time step $\Delta t=10^{-5}$ and total simulation time $t_{\rm sim}= 10^6$.}
\label{fig:tvsr10neg5}
\end{figure}

Figure~\ref{fig:tvsr10neg5} shows an excellent agreement between Eq.~\eqref{eq:mfpt} and numerical Langevin-dynamics simulations of the model, for different parameter values. 
For positive values of $v$, the mean survival time decreases monotonously with the therapy resetting rate, hence it is beneficial to switch slowly among beneficial  therapies  (i.e. keeping  $r$ small) in order to maximize the survival time. For small and even negative values of the efficacy drift, the mean first-passage time is non monotonous; the minimum average survival time takes place for intermediate values of~$r$. For therapies with large negative bias $v<0$, a case that is is relevant in the context of viral evolution, $\langle\tau_r\rangle$ increases monotonically with $r$, i.e. the maximum average survival is achieved switching the therapies as frequent as possible.

\begin{figure}[t]
\includegraphics[width=0.75\linewidth]{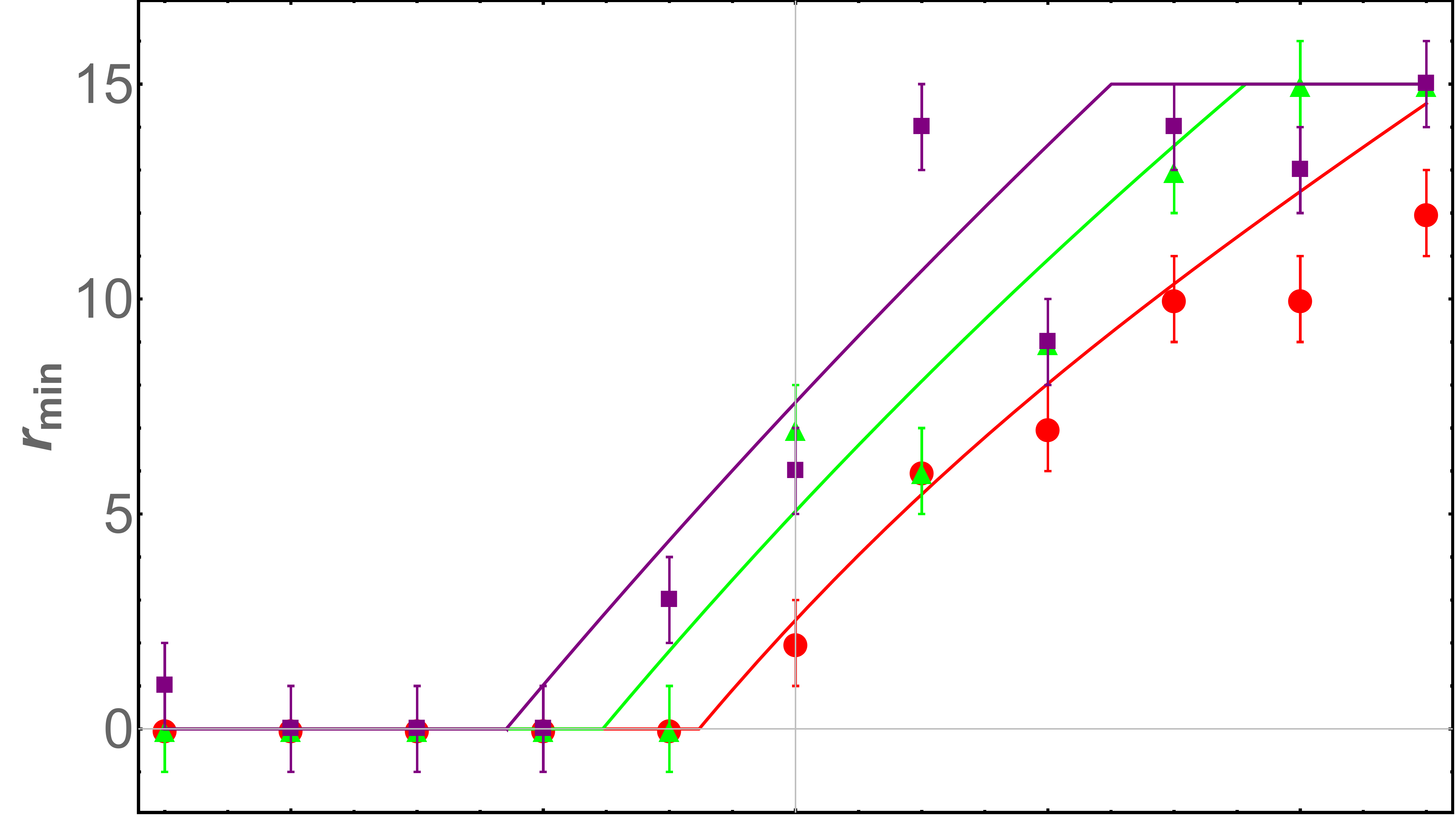}\\
\includegraphics[width=0.75\linewidth]{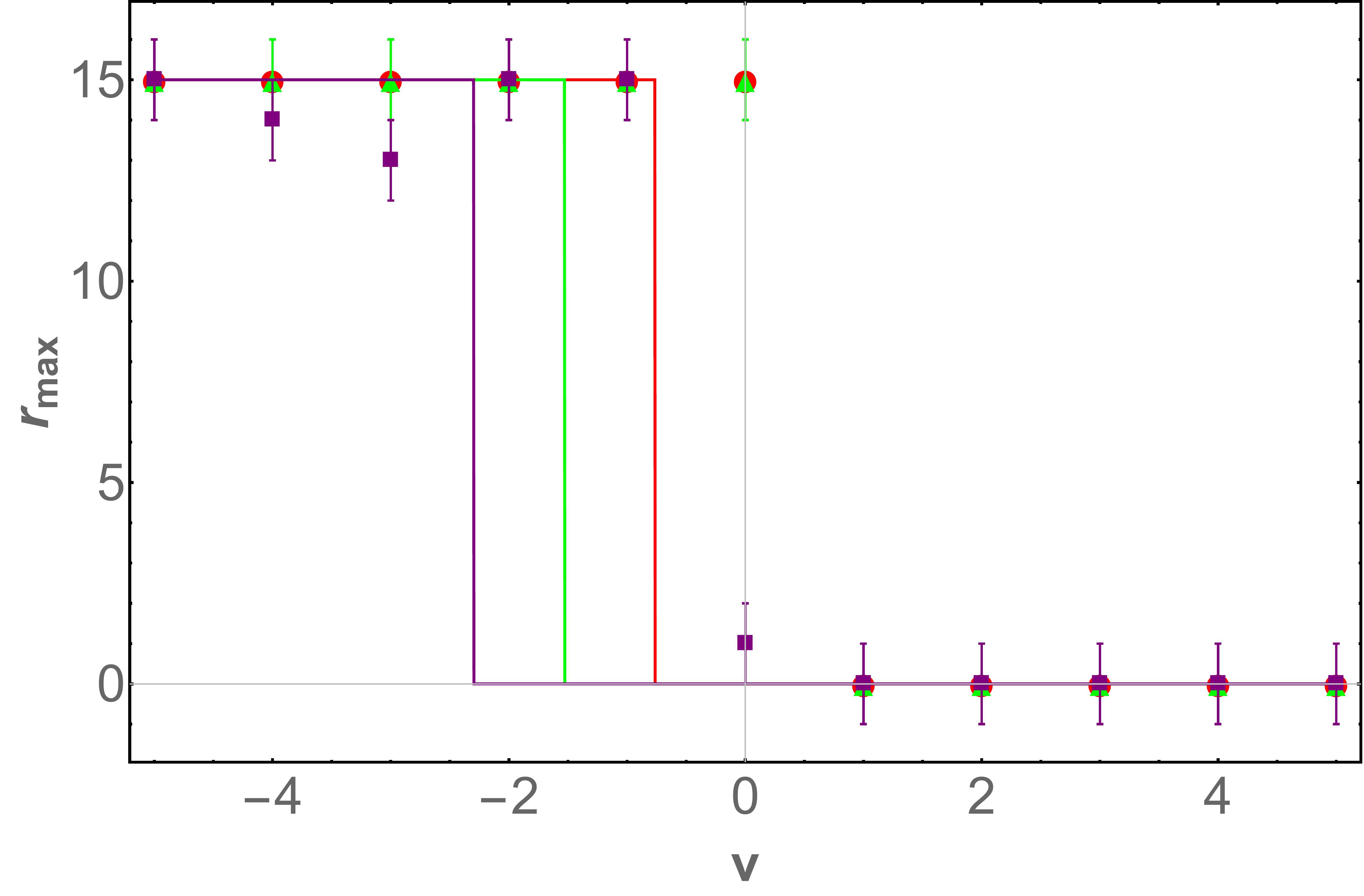}
\caption{
Comparison between simulations (symbols) and   theoretical  results (lines)  of  optimal resetting rates as a function of the therapy efficacy drift:     $r_{\rm min}$ (top) and $r_{\rm max}$ (bottom) denote  the value of the resetting rate at which the mean survival time attains its minimum and maximum values, respectively.  The different curves and symbols are obtained by imposing a maximum allowed resetting rate of $15$,  for different values of $v$ and $D=0.5$ ({\large \color{red}{$\bullet$}}, red line), $D=1$ ({\color{green}{$\blacktriangle$}}, green line) and $D=1.5$ ({\color{violet}{$\blacksquare$}}, purple line). The rest of the simulation parameters are set to  the same values as in Fig.~\ref{fig:tvsr10neg5}.}
\label{fig:3}
\end{figure}

In a real-world scenario, practitioners have to  deal with limited resources such as a finite number of therapies during the life of a patient.  Within this scenario, it is important to know  what is the optimal resetting rate that achieves a desired value of the mean survival time of the patient. To study this problem, we evaluate in Fig.~\ref{fig:3} the resetting rates $r_{\rm min}$  and $r_{\rm max}$ for which the mean survival time attains respectively its minimum and maximum value, within a finite range of resetting rate. We find that $r_{\rm min}=0$ for rapidly evolving virus ($v$ large and negative), whereas $r_{\rm min}$ increases monotonously with $v$ for larger values. Notably $r_{\rm min}$ presents a non-analyic behaviour at a critical value of $v$, and hence of the Peclet number, as reported by recent work on drift-diffusion processes with one absorbing boundary~\cite{ray2019pecletdriftdiffusion}. On the other hand, our results show that the "optimal" resetting rate $r_{\rm max}$ achieving the maximum mean survival time displays a dependency with the therapy drift that has reminiscences of a first-order phase transition;  it exhibits a sudden jump from the maximum allowed resetting rate (for rapidly evolving virus) to zero. Such transition occurs at a critical value of $v$ that depends on the fluctuations $D$ of the therapy efficacy; it can take place at biologically relevant values ($v$ negative) when $D$ is large enough. 



\begin{figure}[t]
\includegraphics[width=0.85\linewidth]{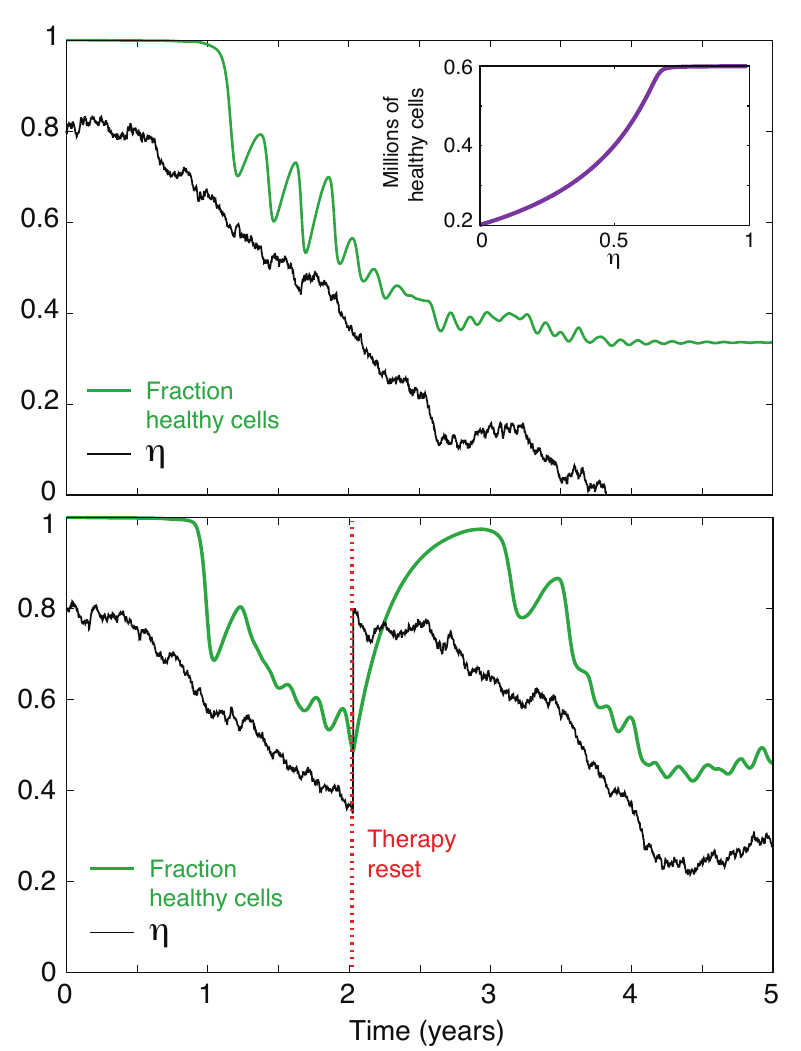}
\caption{Numerical simulations of the viral evolution population-dynamics model given by Eqs.~(\ref{eq:6}-\ref{eq:9}) in the absence (top) and in the presence (bottom) of therapy resetting: fraction of healthy cells with respect to the initial value $H/H_0$ (green line) and therapy efficacy (black line) as a function of time. The inset in the top panel shows the average number of healthy cells as a function of the therapy efficacy in the absence of resetting, and the red dashed line in the bottom panel illustrates the time at which the antiviral therapy efficacy is restored to its initial value. Parameters of the simulations: $v=-3\times 10^{-4}\text{days}^{-1}$ and $D=10^{-5}\text{ days}^{-1}$, $\alpha=6000 \text{ days}^{-1}\text{ ml}^{-1}, \lambda_H = 0.01 \text{ days}^{-1}, \beta=5 \times 10^{-6} \text{ml  days}^{-1}, \epsilon=0.01, p=0.2 \text{ days}^{-1}, a_L=0.1 \text{ days}^{-1}, \lambda_L=0.01 \text{ days}^{-1}, \lambda_I=1 \text{ days}^{-1}$, $K=100\text{ cells ml}^{-1}$, with $r=0$ (top) and $r=(1/3)\text{years}^{-1}$ (bottom), with initial condition $\eta_0=0.8$, $H_0=6\times 10^5 \text{ cells ml}^{-1}, L_0=1\text{ cells ml}^{-1}, I_0=0\text{ cells ml}^{-1}$, and step size $\Delta t=1 \text{days}$. The inset (up) shows the value of $H$ at the non-trivial fixed point of the system Eqs.~(\ref{eq:6}-\ref{eq:8}) for a fixed value of $\eta$.}
\label{fig:4}
\end{figure}

We now consider a stochastic mean-field population-dynamics model describing a multicellular  organism containing healthy $H$, latent $L$, and productively infective $I$ cells by a chronic viral disease such as HIV-1. 
The state of the system is described by its time-dependent numbers $H$, $I$ and $L$, whose dynamics is driven by the stochastic-resetting therapy efficacy $\eta$. 
We remark that we include in the model a population of latent cells to account for  a chronic infection, inspired in previous   mathematical models of HIV-1 \cite{perelsonnelsonhiv-1}.   The dynamics of the model is given by three coupled ordinary differential equations [Eqs.~(\ref{eq:6}-\ref{eq:8}) below] driven by an autonomous stochastic differential equation [Eq.~(\ref{eq:9}) below]:
\begin{eqnarray}
  \dfrac{\text{d}H}{\text{d}t} & = & \alpha - \lambda_H H - (1-\eta) \beta HI \label{eq:6} \\
  \dfrac{\text{d}L}{\text{d}t} & = & \epsilon(1-\eta) \beta HI + pL\left(1-\dfrac{L}{K}\right) - a_L L- \lambda_L L \label{eq:7}\\
  \dfrac{\text{d}I}{\text{d}t} & = & (1-\epsilon) (1-\eta)\beta HI + a_L L - \lambda_I I \label{eq:8}\\
  \text{d}\eta & = & (1-\chi)( v\text{d}t + \sqrt{2D}\text{d}W) + \chi(\eta_0-\eta ).\label{eq:9}
\end{eqnarray}
Here, $\alpha$ denotes the rate of recruitment of new healthy cells, $\lambda_H$ is the death rate of healthy cells, $\beta$ is the infection rate, $\epsilon$ is the probability of an infection resulting into a latent cell, $p$ is the proliferation rate of latent infected cells, $K$ is the carrying capacity which introduces a logistic growth, $a_L$ the activation rate of latent cells, $\lambda_L$ death rate of latent cells, and $\lambda_I$ the death rate of infected cells. Note that the infection rate $\beta$ is multiplied by   the instantaneous probability $(1-\eta)$ of the infection to occur.
Finally, $\chi$ is a binary variable which equals to one (zero) when a reset occurs (does not occur) with probability $r\text{d}t$ ($1-r\text{d}t$), and $W$ is the Wiener process. Hence, Eq.~\eqref{eq:9} complemented with mixed absorbing boundary conditions describes the previously introduced   therapy efficacy stochastic model.
Further details of the model can be found in the Supplemental Material.

\begin{figure}[t]
\includegraphics[width=1.05\linewidth]{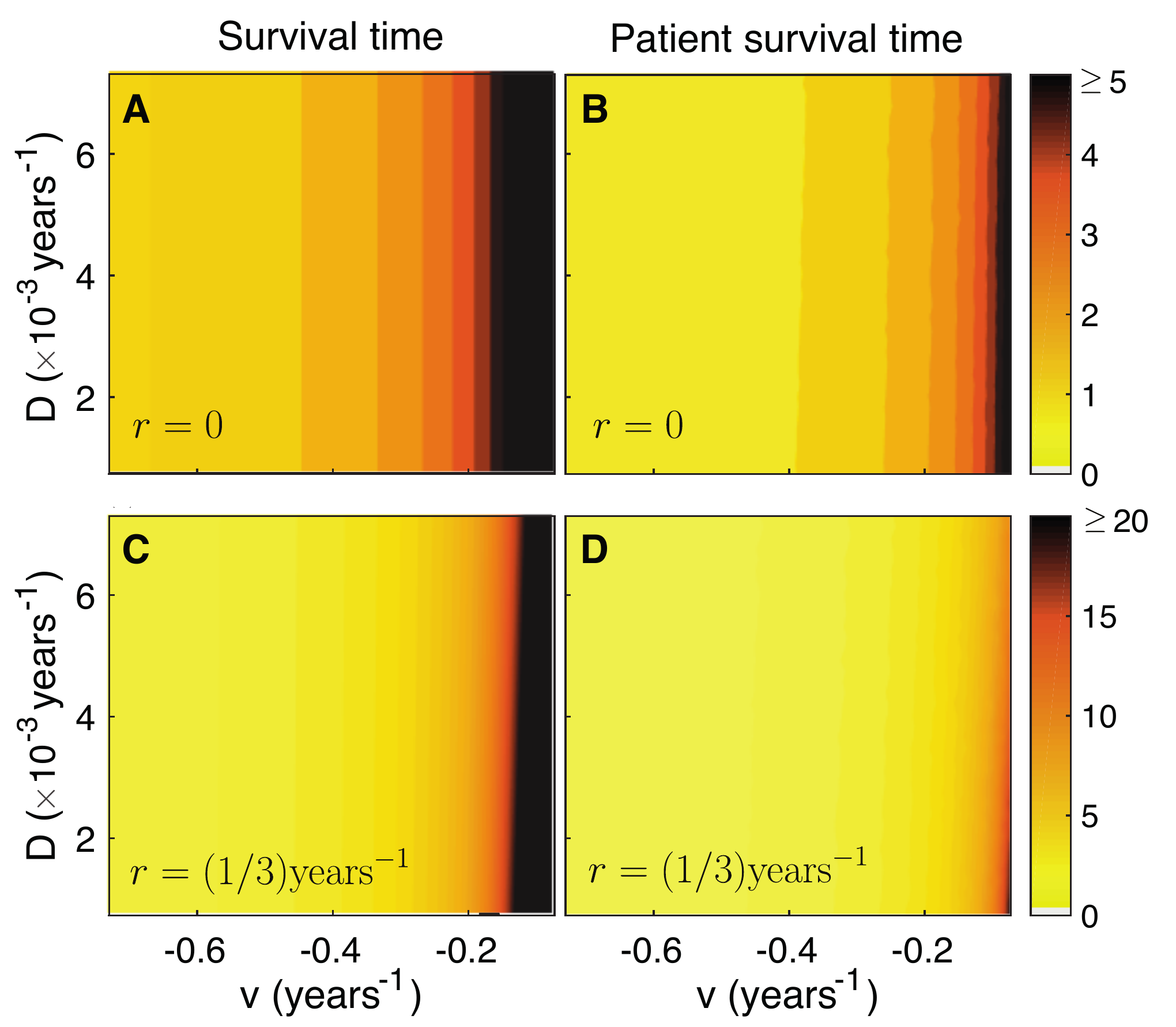}
\caption{Mean survival  times (in years) as a function of the therapy efficacy drift $v$ and diffusion $D$. (A,C) Analytical value of the mean first-passage time for the therapy efficacy $\eta$ to reach the absorbing boundary $\eta=0$,  given by Eq.~\eqref{eq:mfpt}. (B,D) Numerical value of the  mean time elapsed until the fraction of healthy cells falls below $1/2$ its initial value (B,D). 
  The color maps show the values obtained for resetting rates $r=0$ (A,B) and $r=(1/3)\text{years}^{-1}$ (C,D). In (B,D) the values of the simulation parameters were set to the same values as in Fig.~\ref{fig:4} except the time step $\Delta t=0.01 \text{days}$, and  the averages are  done over $2000$ numerical simulations of Eqs.~(\ref{eq:6}-\ref{eq:9}). 
\label{fig:heatmapparameters}}
\end{figure}

Next, we illustrate the model 
with numerical simulations 
showing the impact  of the therapy efficacy in the number of healthy cells of a patient. For this purpose we numerically integrate Eqs.~(\ref{eq:6}-\ref{eq:9}). In this case, we only consider therapies with $v<0$ because mutations beneficial to the virus are dominant in its  evolution.
When changes in therapy are not allowed we observe a drift in $\eta$ through the absorbing barrier, followed by the healthy cells (Fig. \ref{fig:4}, top). Under stochastic changes in the therapy  $\eta$ still drifts through the absorbing state barrier, however, after the stochastic reset, we often observe a period in which the healthy cells recover, delaying the absorption time of $\eta$ (Fig. \ref{fig:4}, bottom). 
We remark that, when $\eta$ reaches the absorbing boundary at zero, the cell population still evolves, towards its fixed point.  Note that, unlike in our first model,  we allow here  for resets from $\eta=0$ to $\eta_0$ because even in the absence of therapy the patient can survive until their therapy  changes. 
This motivates us to study first-passage times in the context of healthy cells.  In doing so, we define the {\em patient survival time} as the time elapsed until  $H\leq \alpha/(2\lambda_H)$, which corresponds to the time until it falls below half of the fixed point in the absence of infection.

We now  determine the impact of changes in the therapy in a  region of parameters for $v$ and $D$.  Figure~\ref{fig:heatmapparameters} shows the mean survival time (left panels) and the mean patient survival time (right panels)  as a function of $v$ and $D$ obtained from  analytical and numerical calculations.
For the parameter values studied here, resetting therapies increases both the mean survival time and the mean patient survival time. Their qualitative behaviour is similar: when the drug resistance develops slowly ($v$ negative but small) the survival time and the patient survival time are large, and vice versa. The larger the resistance fluctuations $D$, the lower the survival times, however this dependency is weaker than for $v$. Notably, when executing therapy resets at a rate of $(1/3)$years$^{-1}$, the mean patient survival time can exceed 20 years for small values of $v$ and $D$ (Fig.~\ref{fig:heatmapparameters}D).

    We have introduced a  multi-scale stochastic model linking  drug resistance development described by a one-dimensional stochastic resetting process with cell population dynamics.      We have found  that the number of healthy cells in a chronic disease such as HIV-1 display negative correlation with the drug resistance. Our analytical and numerical results quantify the beneficial aspects of therapy changes at random Poissonian times for the mean survival time of a patient as a function of the viral evolution  parameters.  We have derived an analytical expression for the mean survival time of the stochastic-resetting therapy efficacy
    as a function of its drift and diffusivity. This expression can be used to estimate patient survival times  in some limits. It will be interesting to extend our work to e.g. compare different therapy resetting protocols under time constraints, account for drug resistance development which depends on the instantaneous viral load,  account for multiple therapies~\cite{richman2009challenge},  etc. We expect potential applications of our work to  determine the epidemiological impact of drug resistance development, using models where the immunity to certain drugs can be transported between individuals.

\renewcommand{\thesection}{S\arabic{section}}
\renewcommand{\thefigure}{S\arabic{figure}}
\renewcommand{\theequation}{S\arabic{equation}}
\renewcommand{\thetable}{S\arabic{table}}

\newpage
\onecolumngrid
\section*{SUPPLEMENTAL MATERIAL}
\appendix

\section{S1. Analytical expression for the mean survival  time} \label{Appendix:1}

In this section we provide additional details about the derivation of Eq.~(5) in the Main Text for the mean absorption time for the one-dimensional stochastic-resetting process that we use to describe the therapy efficacy. The derivation proceeds as follows:
First, we derive an analytical expression for the survival probability in the absence of resetting $r=0$.
    Next, we use a known relation to  derive the Laplace transform of survival probability with resetting from the survival probability without resetting.
    Third, we derive the mean  survival time from the Laplace transform of the survival probability with resetting.

\subsection{First passage   without resetting}

We first consider for the ease of analytical calculations the case $r=0$ in which the therapy efficacy is a drift diffusion process. We denote by  $P (\eta, t)$  the conditional probability density that the efficacy of the therapy is $\eta$ at time $t$, given that its initial value at time $t=0$ was $\eta_0$. It obeys the Fokker-Planck equation that results from taking $r=0$ in Eq.~(1) in the Main Text
\begin{equation}
\dfrac{\partial P (\eta, t)}{\partial t} + \dfrac{\partial J(\eta,t)}{\partial \eta}  = 0,
\label{eq:FPEr0}
\end{equation}
where  $J(\eta,t)$ is the probability current defined in this case as
\begin{equation}
    J(\eta,t)\equiv v P (\eta, t) - D \dfrac{\partial P (\eta, t)}{\partial \eta}.
    \label{eqn:app:J}
    \end{equation}
    Equations~\eqref{eq:FPEr0} and~\eqref{eqn:app:J}   are complemented with the  initial and boundary conditions of the process
\begin{eqnarray}
P (\eta,  t=0) &=& \delta (\eta - \eta_0), \label{eq:S12_1}\\ 
\displaystyle \lim_{\eta \rightarrow 0^+} P (\eta, t) &=& 0,\label{eq:S12_2}\\ 
J (\eta = 1, t) &=& 0.\label{eq:S12_3} 
\label{eqn:app:bc}
\end{eqnarray}
Here, Eq.~\eqref{eq:S12_1} accounts for the initial condition $\eta(0) = \eta_0$, Eq.~\eqref{eq:S12_2} for the absorbing boundary at $\eta = 0$, and Eq.~\eqref{eq:S12_3} for the reflecting boundary condition at $\eta = 1$. 
Taking  the Laplace transform on Eq.~\eqref{eq:FPEr0}, we get
\begin{equation}
    s \mathcal{P} (\eta, s) - P (\eta, t = 0) + v \mathcal{P}' (\eta, s) - D \mathcal{P}'' (\eta, s) = 0,
    \label{eq:laplace0}
\end{equation}
where the prime denotes derivative with respect to $\eta$ and we have introduced the notation 
\begin{equation}
\mathcal{P} (\eta, s) \equiv \int_0^\infty e^{- s t} P (\eta, t) \text{ d} t,
\end{equation}
for the Laplace transform of the propagator.
Applying the initial condition~\eqref{eq:S12_1} into Eq.~\eqref{eq:laplace0}, we obtain
\begin{equation}
s \mathcal{P} (\eta, s) - \delta (\eta - \eta_0) + v \mathcal{P}' (\eta, s) - D \mathcal{P}'' (\eta, s) = 0.
\label{eq:laplace1}
\end{equation}

We now concentrate our efforts in obtaining an analytical solution to Eq.~\eqref{eq:laplace1}. 
First we express $\mathcal{P} (\eta, s)$ as a piecewise continuous function:
\begin{gather}
\mathcal{P} (\eta, s) =
\left\{
\begin{array}{lcl}
\mathcal{P}_< (\eta, s) & \text{ when } & \eta < \eta_0\vspace{5pt}\\
\mathcal{P}_> (\eta, s) & \text{ when } & \eta > \eta_0,
\label{eqn:app:pseparate}
\end{array}
\right.
\end{gather}
for $\eta \ne \eta_0$. Both $\mathcal{P}_< $ and $\mathcal{P}_> $ obey Eq.~\eqref{eq:laplace1} in $\eta<\eta_0$ and $\eta>\eta_0$ respectively, i.e.
\begin{eqnarray}
D \mathcal{P}_<'' (\eta, s)  - v \mathcal{P}_<' (\eta, s) - s \mathcal{P}_< (\eta, s) &=& 0, \label{eq:s19}\\
D \mathcal{P}_<'' (\eta, s)  - v \mathcal{P}_<' (\eta, s) - s \mathcal{P}_< (\eta, s) &=& 0.
\label{eq:s20}
\end{eqnarray}
 The  solutions  of Eqs.~(\ref{eq:s19}-\ref{eq:s20}) are given by  sums of  exponential functions
\begin{eqnarray}
\mathcal{P}_< (\eta, s) &=& a_+ e^{\alpha_+ \eta} + a_- e^{\alpha_- \eta},\label{eq:s21}\\
\mathcal{P}_> (\eta, s) &=& b_+ e^{\alpha_+ \eta} + b_- e^{\alpha_- \eta}.\label{eq:s22}
\end{eqnarray}
Plugging in Eq.~\eqref{eq:s21} in~\eqref{eq:s19} and Eq.~\eqref{eq:s22} in~\eqref{eq:s20} yields the second order equation
\begin{equation}
D \alpha_{\pm}^2 - v \alpha_{\pm} + s = 0,
\end{equation}
whose solutions are given by
\begin{equation}
    \alpha_\pm =\dfrac{v}{2 D} \pm w,  \label{eq:s24}
\end{equation}
where
\begin{equation}
    w \equiv \dfrac{\sqrt{v^2 + 4 D s}}{2 D}. \label{eqn:app:w}
\end{equation}
The parameters  $a_+$, $a_-$, $b_+$, and $b_-$ can be determined using the boundary conditions. 
Applying the absorbing boundary condition~\eqref{eq:S12_2}  into Eq.~\eqref{eq:s21} we get $a_+ + a_- = 0$,  hence defining $a\equiv a_+$, we find that $\mathcal{P}_<$ has the form
\begin{equation}
\mathcal{P}_< (\eta, s) = a \Big[  e^{\alpha_+ \eta} - e^{\alpha_- \eta} \Big].\label{eq:s26}
\end{equation}
Next, we use the reflecting boundary condition~\eqref{eqn:app:bc} in \eqref{eq:s22} which implies $
v \mathcal{P}_> (\eta = 1, s) = D \mathcal{P}_>' (\eta = 1, s)$. This results in the  following relation $v \Big[ b_+ e^{\alpha_+} + b_- e^{\alpha_-} \Big] = D \Big[ \alpha_+ b_+ e^{\alpha_+} + \alpha_- b_- e^{\alpha_-} \Big]$, or equivalently $(v - D \alpha_+) e^{\alpha_+} b_+ = (D \alpha_- - v) e^{\alpha_-} b_- \equiv b$.
Thus we get
\begin{equation}
\mathcal{P}_> (\eta, s) = b \left[ \dfrac{e^{\alpha_+ (\eta - 1)}}{v - D \alpha_+} - \dfrac{e^{\alpha_- (\eta - 1)}}{v - D \alpha_-} \right].
\label{eq:s27}
\end{equation}
The continuity condition of $\mathcal{P}(\eta,s)$ at $\eta = \eta_0$ in Eqs.~(\ref{eq:s26}-\ref{eq:s27}) implies that
\begin{gather}
b \left[ \dfrac{e^{\alpha_+ (\eta_0 - 1)}}{v - D \alpha_+} - \dfrac{e^{\alpha_- (\eta_0 - 1)}}{v - D \alpha_-} \right] = a \left[  e^{\alpha_+ \eta_0} - e^{\alpha_- \eta_0} \right] \equiv c
\end{gather}
therefore
\begin{eqnarray}
\mathcal{P}_< (\eta, s) &=& c \left[ \dfrac{\hspace{5pt}e^{\alpha_+ \eta} - e^{\alpha_- \eta}\hspace{5pt}}{\hspace{5pt}e^{\alpha_+ \eta_0} - e^{\alpha_- \eta_0}\hspace{5pt}} \right]\vphantom{\dfrac{\dfrac{1}{1}}{\dfrac{1}{1}}},\label{eq:s29}\\
\mathcal{P}_> (\eta, s) &=& c \left[ \dfrac{\hspace{5pt}\dfrac{e^{\alpha_+ (\eta - 1)}}{v - D \alpha_+} - \dfrac{e^{\alpha_- (\eta - 1)}}{v - D \alpha_-}\hspace{5pt}}{\hspace{5pt}\dfrac{e^{\alpha_+ (\eta_0 - 1)}}{v - D \alpha_+} - \dfrac{e^{\alpha_- (\eta_0 - 1)}}{v - D \alpha_-}\hspace{5pt}} \right].\label{eq:s30}
\end{eqnarray}
To solve for $c$, we solve the differential equation~\eqref{eq:laplace1} at $\eta = \eta_0$,
\begin{gather}
D \mathcal{P}'' (\eta, s) = v \mathcal{P}' (\eta, s) + s \mathcal{P} (\eta, s) - \delta (\eta - \eta_0).
\end{gather}
Take the integral both sides with respect to  $\eta$ from $\eta_0 - \epsilon$ to $\eta_0 + \epsilon$ and then  the limit of $\epsilon$ to 0 we obtain
\begin{gather}
D \Big[ \mathcal{P}_>' (\eta_0, s) - \mathcal{P}_<' (\eta_0, s) \Big] = v \Big[ \mathcal{P}_> (\eta_0, s) - \mathcal{P}_< (\eta_0, s) \Big]  - 1
\end{gather}
The first term at the right-hand side vanishes due to the continuity condition, therefore
\begin{gather}
\mathcal{P}_>' (\eta_0, s) - \mathcal{P}_<' (\eta_0, s) = - \dfrac{1}{D}. \label{eq:s33}
\end{gather}
Using Eqs.~\eqref{eq:s29} and~\eqref{eq:s30} in~\eqref{eq:s33} and solving for $c$ we get, after some cumbersome simplifications,
\begin{gather}
c = \dfrac{2 \left( \sqrt{v^2 + 4 D s} \cosh \left[ \dfrac{\sqrt{v^2 + 4 D s} (\eta_0 - 1)}{2 D} \right] + v \sinh \left[ \dfrac{\sqrt{v^2 + 4 D s} (\eta_0 - 1)}{2 D} \right] \right) \sinh \left[ \dfrac{\sqrt{v^2 + 4 D s} \eta_0}{2 D} \right]}{(v^2 + 4 D s) \cosh \left[ \dfrac{\sqrt{v^2 + 4 D s}}{2 D} \right] - v \sqrt{v^2 + 4 D s} \sinh \left[ \dfrac{\sqrt{v^2 + 4 D s}}{2 D} \right]}.\label{eq:s34}
\end{gather}

The Laplace transform of the first-passage (survival)  time probability at $\eta = 0$ without resetting  is given by~\cite{redner2001guide}
\begin{gather}
\mathcal{F}_0 (s) = D {\left. \dfrac{\partial}{\partial \eta} \mathcal{P}_< (\eta, s) \right|}_{\eta = 0} - {v \mathcal{P}_< (\eta, s) \bigg|}_{\eta = 0},
\label{eq:s35}
\end{gather}
where the second term at the right-hand side vanishes because of the absorbing boundary condition.  Note that here we have introduced the subscript $0$ to emphasize that this first-passage statistic refers to the case $r=0$. 
Since $c$ does not depend on $\eta$, Eq.~\eqref{eq:s35} together with Eq.~\eqref{eq:s29} imply that 
\begin{gather}
\mathcal{F}_0 (s) = cD  \dfrac{\alpha_+ - \alpha_-}{e^{\alpha_+ \eta_0} - e^{\alpha_- \eta_0}}.
\label{eq:s36}
\end{gather}
Substituing the values of $\alpha_+$ and $\alpha_-$ [Eq.~\eqref{eq:s24}] and $C$ [Eq.~\eqref{eq:s34}] in Eq.~\eqref{eq:s36} we arrive at the analytical expression for the Laplace transform of the first-passage-time probability for the drift-diffusion process with mixed boundary conditions:
\begin{gather}
\mathcal{F}_0 (s) = \dfrac{\sqrt{v^2 + 4 D s} \left(\sqrt{v^2 + 4 D s} \cosh \left[ \dfrac{\sqrt{v^2 + 4 D s} (\eta_0 - 1)}{2 D} \right] + v \sinh \left[ \dfrac{\sqrt{v^2 + 4 D s} (\eta_0 - 1)}{2 D} \right] \right)}{\exp \left[ \dfrac{\eta_0 v}{2 D} \right] \left( (v^2 + 4 D s) \cosh \left[ \dfrac{\sqrt{v^2 + 4 D s}}{2 D} \right] - v \sqrt{v^2 + 4 D s} \sinh \left[ \dfrac{\sqrt{v^2 + 4 D s}}{2 D} \right] \right)}.
\label{eq:s37}
\end{gather}

From Eq.~\eqref{eq:s37} we can also derive an analytical expression for the Laplace transform of the survival probability, which is defined as follows
\begin{eqnarray}
\mathcal{S}_0 (s) &\equiv & \int_{0^+}^1 \text{ d} \eta \mathcal{P} (\eta, s)\\
&=&\int_{0^+}^{\eta_0} \mathcal{P}_< (\eta, s) \text{ d} \eta + \int_{\eta_0}^1 \mathcal{P}_> (\eta, s) \text{ d} \eta,
\end{eqnarray}
where in the second line we have used Eq.~\eqref{eqn:app:pseparate}.
We now recall the relation (valid for any $\eta_0>0$) between the survival probability and the first passage probability $F_0 (t) = - \partial_t S_0 (t)$ which takes the form $\mathcal{F}_0 (s) = - \left[ s \mathcal{S}_0 (s) - S_0 (t = 0) \right]$  upon taking the Laplace transform, or equivalently
\begin{gather}
\mathcal{F}_0 (s) = 1 - s \mathcal{S}_0 (s).
\label{eqn:app:lapfonq0}
\end{gather}
Here we haved used the fact that at time $t = 0$, the process  has not been absorbed by the absorbing boundary $\eta = 0$, i.e. $\eta_0>0$.

\subsection{First passage with resetting}
 
Following~\cite{roldan2017path,pal2019firstpassageresettinginterval}, the survival probability with resetting (denoted here with subscript $r$) obeys a renewal equation together with the survival probability without resetting  
\begin{gather}
S_r ( t) = e^{- r t} S_0 ( t) + r \int_0^t \text{ d} \tau e^{- r \tau} S_0 ( \tau) S_r ( t - \tau) \label{eq:s41}
\end{gather}
Taking the Laplace transform on both sides of Eq.~\eqref{eq:s41}, we  get
$\mathcal{S}_r (s) = \mathcal{S}_0 (s + r) + r \mathcal{S}_0 (s + r) \mathcal{S}_r (s)$, which solving for $\mathcal{S}_r (s)$ yields
\begin{equation}
\mathcal{S}_r (s) = \dfrac{\mathcal{S}_0 (s + r)}{1 - r \mathcal{S}_0 (s + r)}.
\label{eq:s44}
\end{equation}
Therefore, the Laplace transform of the first-passage-time probability with resetting can be expressed in terms of the Laplace transform of the survival probability without resetting as follows
\begin{gather}
\mathcal{F}_r (s) = 1 - \dfrac{s \mathcal{S}_0 (s + r)}{1 - r \mathcal{S}_0 (s + r)}.
\label{eq:s43}
\end{gather}
Shifting the relation~\eqref{eqn:app:lapfonq0} from $s$ to $s+r$, i.e. using $\mathcal{S}_0 (s + r) = \left(1 - \mathcal{F}_0 (s + r)\right)/(s + r)$ in Eq.~\eqref{eq:s43} we obtain after some simplifications
\begin{gather}
\mathcal{F}_r (s) = \dfrac{\Big( r + s \Big) \mathcal{F}_0 (s + r)}{s + r \mathcal{F}_0 (s + r)} \label{eqn:app:frs}
\end{gather}
After some algebra, in using Eq.~\eqref{eq:s37} in~\eqref{eqn:app:frs} and the relation $\langle\tau_r\rangle = -\lim_{s\to 0} \partial_s \mathcal{F}_r (s) $ we derive Eq.~(5) in the Main Text for the mean first-passage time, copied here for convenience 

\begin{gather}
\big< \tau_r \big> = 
\dfrac{
\begin{array}{r}
{\text{Pe}}^2 \left(
\begin{array}{l}
\exp \Big[ \text{Pe}\, \eta_0 \Big] \cosh \Big[ \Omega \big( \eta_0 - 2 \big) \Big] \vspace{5pt}\\
- \cosh \Big[ 2 \Omega \big( \eta_0 - 1 \big) \Big]
\end{array}
\right) \vspace{5pt}\\
+ \text{Pe}\, \Omega \left(
\begin{array}{l}
\exp \Big[ \text{Pe}\, \eta_0 \Big] \sinh \Big[ \Omega \big( \eta_0 - 2 \big) \Big] \vspace{5pt}\\
- \sinh \Big[ 2 \Omega \big( \eta_0 - 1 \big) \Big]
\end{array}
\right)
\end{array}
- \xi \left(
\begin{array}{l}
1 \vphantom{\dfrac{1}{1}}\vspace{5pt}\\
+ \cosh \Big[ 2 \Omega \big( \eta_0 - 1 \big) \Big] \vspace{5pt}\\
- \exp \Big[ \text{Pe}\, \eta_0 \Big] \cosh \Big[ \Omega \eta_0 \Big] \vspace{5pt}\\
- \exp \Big[ \text{Pe}\, \eta_0 \Big] \cosh \Big[ \Omega \big( \eta_0 - 2 \big) \Big]
\end{array}
\right)
}{
r {\left\{
\text{Pe} \sinh \Big[ \Omega \big( \eta_0 - 1 \big) \Big] 
+ \Omega \cosh \Big[ \Omega \big( \eta_0 - 1 \big) \Big]
\right\}}^2
}
\end{gather}
where we have introduced the variables
\begin{equation}
    \Omega = \dfrac{\sqrt{v^2 + 4 D r}}{2 D};\quad
    \text{Pe} = \dfrac{v}{2 D};\quad
    \xi = \dfrac{r}{2 D}.
\end{equation}

\section{S2. Details of the biophysical model}\label{ape:modeldetails}

Our biological model [Eqs. (6-9) in the Main Text] is an adaptation of that  reported in Ref.~\cite{perelsonnelsonhiv-1} and describes the evolution of three different cell populations in blood: healthy cells $H$ which are susceptible to be infected, infected cells $L$ which are in a latent state, and productively infected cells $I$ which can infect healthy cells. The dynamics is described by  the reactions described below, where the chronic feature is modelled by adding a latent population with a logistic growth. Such dynamics have been hypothesized in \cite{pierson2000reservoirs} and used to model  HIV-1 dynamics in~\cite{perelsonnelsonhiv-1, sanchez-taltavull2016hiv-1eradication}. 

\begin{itemize}

\item Recruitment of  a new healthy cell by the organism at a rate $a$: $\emptyset \to H$.

\item Death of a healthy at a rate $\lambda_H$: $H \to \emptyset$.

\item A healthy cell being infected, which we assume it is proportional to the number of infected cells, $I$, and inversely proportional to the efficacy of the therapy $\eta$, at a rate $\beta$. With probability $\epsilon$ the resulting cell will be a latent cell: $H \to L$, and with probability $1-\epsilon$ the resulting cell will be a productively infected cell: $H \to I$.

\item Proliferation of a latent cell with rate $p$: $L \to L+L$.

\item Death of latent cells by competition for resources, with rate $\frac{p}{K}$: $L+L \to \emptyset$. This is an artificial reaction to mimic the logistic growth, see \cite{sanchez-taltavull2016hiv-1eradication} for details in HIV-1 modeling.

\item Activation of a latent cell resulting into a productively infected cell at rate $a_L$: $L\to I$.

\item Death of a latent cell at rate $\lambda_L$: $L \to \emptyset$.

\item Death of a productively infected cell at rate $\lambda_I$: $I \to \emptyset$.

\end{itemize}
A mean-field formulation of the dynamics of the system leads to the ordinary differential equations (6-8) in the Main Text.









\begin{thebibliography}{38}
\expandafter\ifx\csname natexlab\endcsname\relax\def\natexlab#1{#1}\fi
\expandafter\ifx\csname bibnamefont\endcsname\relax
  \def\bibnamefont#1{#1}\fi
\expandafter\ifx\csname bibfnamefont\endcsname\relax
  \def\bibfnamefont#1{#1}\fi
\expandafter\ifx\csname citenamefont\endcsname\relax
  \def\citenamefont#1{#1}\fi
\expandafter\ifx\csname url\endcsname\relax
  \def\url#1{\texttt{#1}}\fi
\expandafter\ifx\csname urlprefix\endcsname\relax\def\urlprefix{URL }\fi
\providecommand{\bibinfo}[2]{#2}
\providecommand{\eprint}[2][]{\url{#2}}

\bibitem[{\citenamefont{Lichterfeld and
  Zachary}(2011)}]{lichterfeld2011treating}
\bibinfo{author}{\bibfnamefont{M.}~\bibnamefont{Lichterfeld}} \bibnamefont{and}
  \bibinfo{author}{\bibfnamefont{K.~C.} \bibnamefont{Zachary}},
  \bibinfo{journal}{Ther. Adv. Chronic Dis.} \textbf{\bibinfo{volume}{2}},
  \bibinfo{pages}{293} (\bibinfo{year}{2011}).

\bibitem[{\citenamefont{Clutter et~al.}(2016)\citenamefont{Clutter, Jordan,
  Bertagnolio, and Shafer}}]{clutter2016hiv}
\bibinfo{author}{\bibfnamefont{D.~S.} \bibnamefont{Clutter}},
  \bibinfo{author}{\bibfnamefont{M.~R.} \bibnamefont{Jordan}},
  \bibinfo{author}{\bibfnamefont{S.}~\bibnamefont{Bertagnolio}},
  \bibnamefont{and} \bibinfo{author}{\bibfnamefont{R.~W.}
  \bibnamefont{Shafer}}, \bibinfo{journal}{Infect. Genet. Evol.}
  \textbf{\bibinfo{volume}{46}}, \bibinfo{pages}{292} (\bibinfo{year}{2016}).

\bibitem[{\citenamefont{Pillay and Zambon}(1998)}]{pillay1998antiviral}
\bibinfo{author}{\bibfnamefont{D.}~\bibnamefont{Pillay}} \bibnamefont{and}
  \bibinfo{author}{\bibfnamefont{M.}~\bibnamefont{Zambon}},
  \bibinfo{journal}{Br. Med. J.} \textbf{\bibinfo{volume}{317}},
  \bibinfo{pages}{660} (\bibinfo{year}{1998}).

\bibitem[{\citenamefont{Perelson and Nelson}(1999)}]{perelsonnelsonhiv-1}
\bibinfo{author}{\bibfnamefont{A.~S.} \bibnamefont{Perelson}} \bibnamefont{and}
  \bibinfo{author}{\bibfnamefont{P.~W.} \bibnamefont{Nelson}},
  \bibinfo{journal}{SIAM Rev.} \textbf{\bibinfo{volume}{41}},
  \bibinfo{pages}{3} (\bibinfo{year}{1999}).

\bibitem[{\citenamefont{Sanchez-Taltavull
  et~al.}(2016)\citenamefont{Sanchez-Taltavull, Vieiro, and
  Alarcon}}]{sanchez-taltavull2016hiv-1eradication}
\bibinfo{author}{\bibfnamefont{D.}~\bibnamefont{Sanchez-Taltavull}},
  \bibinfo{author}{\bibfnamefont{A.}~\bibnamefont{Vieiro}}, \bibnamefont{and}
  \bibinfo{author}{\bibfnamefont{T.}~\bibnamefont{Alarcon}},
  \bibinfo{journal}{J. Math. Biol.} \textbf{\bibinfo{volume}{73}},
  \bibinfo{pages}{919} (\bibinfo{year}{2016}).

\bibitem[{\citenamefont{Strasfeld and Chou}(2010)}]{strasfeld2010antiviral}
\bibinfo{author}{\bibfnamefont{L.}~\bibnamefont{Strasfeld}} \bibnamefont{and}
  \bibinfo{author}{\bibfnamefont{S.}~\bibnamefont{Chou}},
  \bibinfo{journal}{Infectious Disease Clinics} \textbf{\bibinfo{volume}{24}},
  \bibinfo{pages}{809} (\bibinfo{year}{2010}).

\bibitem[{\citenamefont{Fabreti et~al.}(2019)\citenamefont{Fabreti, Castro,
  Gorzoni, Janini, and Antoneli}}]{fabreti2019stochastic}
\bibinfo{author}{\bibfnamefont{L.~G.} \bibnamefont{Fabreti}},
  \bibinfo{author}{\bibfnamefont{D.}~\bibnamefont{Castro}},
  \bibinfo{author}{\bibfnamefont{B.}~\bibnamefont{Gorzoni}},
  \bibinfo{author}{\bibfnamefont{L.~M.~R.} \bibnamefont{Janini}},
  \bibnamefont{and} \bibinfo{author}{\bibfnamefont{F.}~\bibnamefont{Antoneli}},
  \bibinfo{journal}{Bulletin of mathematical biology}
  \textbf{\bibinfo{volume}{81}}, \bibinfo{pages}{1031} (\bibinfo{year}{2019}).

\bibitem[{\citenamefont{Manrubia and Lázaro}(2006)}]{manrubiaviralevolution}
\bibinfo{author}{\bibfnamefont{S.~C.} \bibnamefont{Manrubia}} \bibnamefont{and}
  \bibinfo{author}{\bibfnamefont{E.}~\bibnamefont{Lázaro}},
  \bibinfo{journal}{Phys. Life Rev.} \textbf{\bibinfo{volume}{3}},
  \bibinfo{pages}{65} (\bibinfo{year}{2006}).

\bibitem[{\citenamefont{Wensing et~al.}(2019)\citenamefont{Wensing, Calvez,
  Ceccherini-Silberstein, Charpentier, G{\"u}nthard, Paredes, Shafer, and
  Richman}}]{wensing2019update}
\bibinfo{author}{\bibfnamefont{A.~M.} \bibnamefont{Wensing}},
  \bibinfo{author}{\bibfnamefont{V.}~\bibnamefont{Calvez}},
  \bibinfo{author}{\bibfnamefont{F.}~\bibnamefont{Ceccherini-Silberstein}},
  \bibinfo{author}{\bibfnamefont{C.}~\bibnamefont{Charpentier}},
  \bibinfo{author}{\bibfnamefont{H.~F.} \bibnamefont{G{\"u}nthard}},
  \bibinfo{author}{\bibfnamefont{R.}~\bibnamefont{Paredes}},
  \bibinfo{author}{\bibfnamefont{R.~W.} \bibnamefont{Shafer}},
  \bibnamefont{and} \bibinfo{author}{\bibfnamefont{D.~D.}
  \bibnamefont{Richman}}, \bibinfo{journal}{Top. Antivir. Med.}
  \textbf{\bibinfo{volume}{27}}, \bibinfo{pages}{111} (\bibinfo{year}{2019}).

\bibitem[{\citenamefont{Tria et~al.}(2005)\citenamefont{Tria, Laessig, Peliti,
  and Franz}}]{tria2005minimal}
\bibinfo{author}{\bibfnamefont{F.}~\bibnamefont{Tria}},
  \bibinfo{author}{\bibfnamefont{M.}~\bibnamefont{Laessig}},
  \bibinfo{author}{\bibfnamefont{L.}~\bibnamefont{Peliti}}, \bibnamefont{and}
  \bibinfo{author}{\bibfnamefont{S.}~\bibnamefont{Franz}}, \bibinfo{journal}{J.
  Stat. Mech.} \textbf{\bibinfo{volume}{2005}}, \bibinfo{pages}{P07008}
  (\bibinfo{year}{2005}).

\bibitem[{\citenamefont{Nelson et~al.}(2006)\citenamefont{Nelson, Simonsen,
  Viboud, Miller, Taylor, St~George, Griesemer, Ghedin, Sengamalay, Spiro
  et~al.}}]{nelson2006stochastic}
\bibinfo{author}{\bibfnamefont{M.~I.} \bibnamefont{Nelson}},
  \bibinfo{author}{\bibfnamefont{L.}~\bibnamefont{Simonsen}},
  \bibinfo{author}{\bibfnamefont{C.}~\bibnamefont{Viboud}},
  \bibinfo{author}{\bibfnamefont{M.~A.} \bibnamefont{Miller}},
  \bibinfo{author}{\bibfnamefont{J.}~\bibnamefont{Taylor}},
  \bibinfo{author}{\bibfnamefont{K.}~\bibnamefont{St~George}},
  \bibinfo{author}{\bibfnamefont{S.~B.} \bibnamefont{Griesemer}},
  \bibinfo{author}{\bibfnamefont{E.}~\bibnamefont{Ghedin}},
  \bibinfo{author}{\bibfnamefont{N.~A.} \bibnamefont{Sengamalay}},
  \bibinfo{author}{\bibfnamefont{D.~J.} \bibnamefont{Spiro}},
  \bibnamefont{et~al.}, \bibinfo{journal}{PLoS Pathog.}
  \textbf{\bibinfo{volume}{2}}, \bibinfo{pages}{e125} (\bibinfo{year}{2006}).

\bibitem[{\citenamefont{Zitzmann and
  Kaderali}(2018)}]{zitzmann2018mathematical}
\bibinfo{author}{\bibfnamefont{C.}~\bibnamefont{Zitzmann}} \bibnamefont{and}
  \bibinfo{author}{\bibfnamefont{L.}~\bibnamefont{Kaderali}},
  \bibinfo{journal}{Front. Microbiol.} \textbf{\bibinfo{volume}{9}},
  \bibinfo{pages}{1546} (\bibinfo{year}{2018}).

\bibitem[{\citenamefont{Tsimring et~al.}(1996)\citenamefont{Tsimring, Levine,
  and Kessler}}]{tsimring1996rna}
\bibinfo{author}{\bibfnamefont{L.~S.} \bibnamefont{Tsimring}},
  \bibinfo{author}{\bibfnamefont{H.}~\bibnamefont{Levine}}, \bibnamefont{and}
  \bibinfo{author}{\bibfnamefont{D.~A.} \bibnamefont{Kessler}},
  \bibinfo{journal}{Phys. Rev. Lett.} \textbf{\bibinfo{volume}{76}},
  \bibinfo{pages}{4440} (\bibinfo{year}{1996}).

\bibitem[{\citenamefont{Novella et~al.}(1995)\citenamefont{Novella, Duarte,
  Elena, Moya, Domingo, and Holland}}]{novella1995exponential}
\bibinfo{author}{\bibfnamefont{I.~S.} \bibnamefont{Novella}},
  \bibinfo{author}{\bibfnamefont{E.~A.} \bibnamefont{Duarte}},
  \bibinfo{author}{\bibfnamefont{S.~F.} \bibnamefont{Elena}},
  \bibinfo{author}{\bibfnamefont{A.}~\bibnamefont{Moya}},
  \bibinfo{author}{\bibfnamefont{E.}~\bibnamefont{Domingo}}, \bibnamefont{and}
  \bibinfo{author}{\bibfnamefont{J.~J.} \bibnamefont{Holland}},
  \bibinfo{journal}{PNAS} \textbf{\bibinfo{volume}{92}}, \bibinfo{pages}{5841}
  (\bibinfo{year}{1995}).

\bibitem[{\citenamefont{Holland et~al.}(1991)\citenamefont{Holland,
  De~La~Torre, Clarke, and Duarte}}]{holland1991quantitation}
\bibinfo{author}{\bibfnamefont{J.~J.} \bibnamefont{Holland}},
  \bibinfo{author}{\bibfnamefont{J.~C.} \bibnamefont{De~La~Torre}},
  \bibinfo{author}{\bibfnamefont{D.}~\bibnamefont{Clarke}}, \bibnamefont{and}
  \bibinfo{author}{\bibfnamefont{E.}~\bibnamefont{Duarte}},
  \bibinfo{journal}{J. Virol.} \textbf{\bibinfo{volume}{65}},
  \bibinfo{pages}{2960} (\bibinfo{year}{1991}).

\bibitem[{\citenamefont{Evans and Majumdar}(2011)}]{evans2011diffusion}
\bibinfo{author}{\bibfnamefont{M.~R.} \bibnamefont{Evans}} \bibnamefont{and}
  \bibinfo{author}{\bibfnamefont{S.~N.} \bibnamefont{Majumdar}},
  \bibinfo{journal}{Phys. Rev. Lett.} \textbf{\bibinfo{volume}{106}},
  \bibinfo{pages}{160601} (\bibinfo{year}{2011}).

\bibitem[{\citenamefont{Evans et~al.}(2020)\citenamefont{Evans, Majumdar, and
  Schehr}}]{evans2020topicalreviewstochasticresetting}
\bibinfo{author}{\bibfnamefont{M.~R.} \bibnamefont{Evans}},
  \bibinfo{author}{\bibfnamefont{S.~N.} \bibnamefont{Majumdar}},
  \bibnamefont{and} \bibinfo{author}{\bibfnamefont{G.}~\bibnamefont{Schehr}},
  \bibinfo{journal}{J. Phys. A} \textbf{\bibinfo{volume}{53}},
  \bibinfo{pages}{193001} (\bibinfo{year}{2020}).

\bibitem[{\citenamefont{Bhat et~al.}(2016)\citenamefont{Bhat, De~Bacco, and
  Redner}}]{bhat2016stochastic}
\bibinfo{author}{\bibfnamefont{U.}~\bibnamefont{Bhat}},
  \bibinfo{author}{\bibfnamefont{C.}~\bibnamefont{De~Bacco}}, \bibnamefont{and}
  \bibinfo{author}{\bibfnamefont{S.}~\bibnamefont{Redner}},
  \bibinfo{journal}{J. Stat. Mech.} \textbf{\bibinfo{volume}{2016}},
  \bibinfo{pages}{083401} (\bibinfo{year}{2016}).

\bibitem[{\citenamefont{Ku{\'s}mierz and
  Gudowska-Nowak}(2015)}]{kusmierz2015optimal}
\bibinfo{author}{\bibfnamefont{{\L}.}~\bibnamefont{Ku{\'s}mierz}}
  \bibnamefont{and}
  \bibinfo{author}{\bibfnamefont{E.}~\bibnamefont{Gudowska-Nowak}},
  \bibinfo{journal}{Phys. Rev. E} \textbf{\bibinfo{volume}{92}},
  \bibinfo{pages}{052127} (\bibinfo{year}{2015}).

\bibitem[{\citenamefont{Pal and
  Prasad}(2019{\natexlab{a}})}]{pal2019firstpassageresettinginterval}
\bibinfo{author}{\bibfnamefont{A.}~\bibnamefont{Pal}} \bibnamefont{and}
  \bibinfo{author}{\bibfnamefont{V.}~\bibnamefont{Prasad}},
  \bibinfo{journal}{Phys. Rev. E} \textbf{\bibinfo{volume}{99}},
  \bibinfo{pages}{032123} (\bibinfo{year}{2019}{\natexlab{a}}).

\bibitem[{\citenamefont{Pal and
  Prasad}(2019{\natexlab{b}})}]{pal2019landauphasetransitionsstochasticresetting}
\bibinfo{author}{\bibfnamefont{A.}~\bibnamefont{Pal}} \bibnamefont{and}
  \bibinfo{author}{\bibfnamefont{V.}~\bibnamefont{Prasad}},
  \bibinfo{journal}{Phys. Rev. Res.} \textbf{\bibinfo{volume}{1}},
  \bibinfo{pages}{032001(R)} (\bibinfo{year}{2019}{\natexlab{b}}).

\bibitem[{\citenamefont{Durang et~al.}(2019)\citenamefont{Durang, Lee, Lizana,
  and Jeon}}]{durang2019firstpassagestochasticresettingboundeddomain}
\bibinfo{author}{\bibfnamefont{X.}~\bibnamefont{Durang}},
  \bibinfo{author}{\bibfnamefont{S.}~\bibnamefont{Lee}},
  \bibinfo{author}{\bibfnamefont{L.}~\bibnamefont{Lizana}}, \bibnamefont{and}
  \bibinfo{author}{\bibfnamefont{J.-y.} \bibnamefont{Jeon}},
  \bibinfo{journal}{J. Phys. A} \textbf{\bibinfo{volume}{52}},
  \bibinfo{pages}{224001} (\bibinfo{year}{2019}).

\bibitem[{\citenamefont{Christou and
  Schadschneider}(2015)}]{christou2015diffusionresettingbounded}
\bibinfo{author}{\bibfnamefont{C.}~\bibnamefont{Christou}} \bibnamefont{and}
  \bibinfo{author}{\bibfnamefont{A.}~\bibnamefont{Schadschneider}},
  \bibinfo{journal}{J. Phys. A} \textbf{\bibinfo{volume}{48}},
  \bibinfo{pages}{285003} (\bibinfo{year}{2015}).

\bibitem[{\citenamefont{Pal and Reuveni}(2017)}]{pal2017firstpassagerestart}
\bibinfo{author}{\bibfnamefont{A.}~\bibnamefont{Pal}} \bibnamefont{and}
  \bibinfo{author}{\bibfnamefont{S.}~\bibnamefont{Reuveni}},
  \bibinfo{journal}{Phys. Rev. Lett.} \textbf{\bibinfo{volume}{118}},
  \bibinfo{pages}{030603} (\bibinfo{year}{2017}).

\bibitem[{\citenamefont{Chechkin and
  Sokolov}(2019)}]{chechkin2018searchresettingunifiedrenewal}
\bibinfo{author}{\bibfnamefont{A.}~\bibnamefont{Chechkin}} \bibnamefont{and}
  \bibinfo{author}{\bibfnamefont{I.}~\bibnamefont{Sokolov}},
  \bibinfo{journal}{Phys. Rev. Lett.} \textbf{\bibinfo{volume}{121}},
  \bibinfo{pages}{042128} (\bibinfo{year}{2019}).

\bibitem[{\citenamefont{Ray et~al.}(2019)\citenamefont{Ray, Mondal, and
  Reuveni}}]{ray2019pecletdriftdiffusion}
\bibinfo{author}{\bibfnamefont{S.}~\bibnamefont{Ray}},
  \bibinfo{author}{\bibfnamefont{D.}~\bibnamefont{Mondal}}, \bibnamefont{and}
  \bibinfo{author}{\bibfnamefont{S.}~\bibnamefont{Reuveni}},
  \bibinfo{journal}{J. Phys. A} \textbf{\bibinfo{volume}{52}},
  \bibinfo{pages}{255002} (\bibinfo{year}{2019}).

\bibitem[{\citenamefont{Basu et~al.}(2019)\citenamefont{Basu, Kundu, and
  Pal}}]{basu2019symmetric}
\bibinfo{author}{\bibfnamefont{U.}~\bibnamefont{Basu}},
  \bibinfo{author}{\bibfnamefont{A.}~\bibnamefont{Kundu}}, \bibnamefont{and}
  \bibinfo{author}{\bibfnamefont{A.}~\bibnamefont{Pal}},
  \bibinfo{journal}{Phys. Rev. E} \textbf{\bibinfo{volume}{100}},
  \bibinfo{pages}{032136} (\bibinfo{year}{2019}).

\bibitem[{\citenamefont{Rold{\'a}n et~al.}(2016)\citenamefont{Rold{\'a}n,
  Lisica, S{\'a}nchez-Taltavull, and Grill}}]{roldan2016stochastic}
\bibinfo{author}{\bibfnamefont{{\'E}.}~\bibnamefont{Rold{\'a}n}},
  \bibinfo{author}{\bibfnamefont{A.}~\bibnamefont{Lisica}},
  \bibinfo{author}{\bibfnamefont{D.}~\bibnamefont{S{\'a}nchez-Taltavull}},
  \bibnamefont{and} \bibinfo{author}{\bibfnamefont{S.~W.} \bibnamefont{Grill}},
  \bibinfo{journal}{Phys. Rev. E} \textbf{\bibinfo{volume}{93}},
  \bibinfo{pages}{062411} (\bibinfo{year}{2016}).

\bibitem[{\citenamefont{Tucci et~al.}(2020)\citenamefont{Tucci, Gambassi,
  Gupta, and Rold{\'a}n}}]{tucci2020controlling}
\bibinfo{author}{\bibfnamefont{G.}~\bibnamefont{Tucci}},
  \bibinfo{author}{\bibfnamefont{A.}~\bibnamefont{Gambassi}},
  \bibinfo{author}{\bibfnamefont{S.}~\bibnamefont{Gupta}}, \bibnamefont{and}
  \bibinfo{author}{\bibfnamefont{{\'E}.}~\bibnamefont{Rold{\'a}n}},
  \bibinfo{journal}{arXiv preprint arXiv:2005.05173}  (\bibinfo{year}{2020}).

\bibitem[{\citenamefont{Mora}(2015)}]{mora2015physical}
\bibinfo{author}{\bibfnamefont{T.}~\bibnamefont{Mora}}, \bibinfo{journal}{Phys.
  Rev. Lett.} \textbf{\bibinfo{volume}{115}}, \bibinfo{pages}{038102}
  (\bibinfo{year}{2015}).

\bibitem[{\citenamefont{Bressloff}(2020)}]{bressloff2020stochastic}
\bibinfo{author}{\bibfnamefont{P.~C.} \bibnamefont{Bressloff}},
  \bibinfo{journal}{Phys. Rev. E} \textbf{\bibinfo{volume}{102}},
  \bibinfo{pages}{022134} (\bibinfo{year}{2020}).

\bibitem[{\citenamefont{Mercado-V{\'a}squez and
  Boyer}(2018)}]{mercado2018lotka}
\bibinfo{author}{\bibfnamefont{G.}~\bibnamefont{Mercado-V{\'a}squez}}
  \bibnamefont{and} \bibinfo{author}{\bibfnamefont{D.}~\bibnamefont{Boyer}},
  \bibinfo{journal}{J. Phys. A} \textbf{\bibinfo{volume}{51}},
  \bibinfo{pages}{405601} (\bibinfo{year}{2018}).

\bibitem[{\citenamefont{da~Silva and Fragoso}(2018)}]{da2018interplay}
\bibinfo{author}{\bibfnamefont{T.~T.} \bibnamefont{da~Silva}} \bibnamefont{and}
  \bibinfo{author}{\bibfnamefont{M.~D.} \bibnamefont{Fragoso}},
  \bibinfo{journal}{J. Phys. A} \textbf{\bibinfo{volume}{51}},
  \bibinfo{pages}{505002} (\bibinfo{year}{2018}).

\bibitem[{\citenamefont{Garc{\'\i}a-Garc{\'\i}a
  et~al.}(2019)\citenamefont{Garc{\'\i}a-Garc{\'\i}a, Genthon, and
  Lacoste}}]{garcia2019linking}
\bibinfo{author}{\bibfnamefont{R.}~\bibnamefont{Garc{\'\i}a-Garc{\'\i}a}},
  \bibinfo{author}{\bibfnamefont{A.}~\bibnamefont{Genthon}}, \bibnamefont{and}
  \bibinfo{author}{\bibfnamefont{D.}~\bibnamefont{Lacoste}},
  \bibinfo{journal}{Phys. Rev. E} \textbf{\bibinfo{volume}{99}},
  \bibinfo{pages}{042413} (\bibinfo{year}{2019}).

\bibitem[{\citenamefont{Richman et~al.}(2009)\citenamefont{Richman, Margolis,
  Delaney, Greene, Hazuda, and Pomerantz}}]{richman2009challenge}
\bibinfo{author}{\bibfnamefont{D.~D.} \bibnamefont{Richman}},
  \bibinfo{author}{\bibfnamefont{D.~M.} \bibnamefont{Margolis}},
  \bibinfo{author}{\bibfnamefont{M.}~\bibnamefont{Delaney}},
  \bibinfo{author}{\bibfnamefont{W.~C.} \bibnamefont{Greene}},
  \bibinfo{author}{\bibfnamefont{D.}~\bibnamefont{Hazuda}}, \bibnamefont{and}
  \bibinfo{author}{\bibfnamefont{R.~J.} \bibnamefont{Pomerantz}},
  \bibinfo{journal}{Science} \textbf{\bibinfo{volume}{323}},
  \bibinfo{pages}{1304} (\bibinfo{year}{2009}).

\bibitem[{\citenamefont{Redner}(2001)}]{redner2001guide}
\bibinfo{author}{\bibfnamefont{S.}~\bibnamefont{Redner}},
  \emph{\bibinfo{title}{A guide to first-passage processes}}
  (\bibinfo{publisher}{Cambridge University Press}, \bibinfo{year}{2001}).

\bibitem[{\citenamefont{Rold{\'a}n and Gupta}(2017)}]{roldan2017path}
\bibinfo{author}{\bibfnamefont{{\'E}.}~\bibnamefont{Rold{\'a}n}}
  \bibnamefont{and} \bibinfo{author}{\bibfnamefont{S.}~\bibnamefont{Gupta}},
  \bibinfo{journal}{Phys. Rev. E} \textbf{\bibinfo{volume}{96}},
  \bibinfo{pages}{022130} (\bibinfo{year}{2017}).

\bibitem[{\citenamefont{Pierson et~al.}(2000)\citenamefont{Pierson, McArthur,
  and Siliciano}}]{pierson2000reservoirs}
\bibinfo{author}{\bibfnamefont{T.}~\bibnamefont{Pierson}},
  \bibinfo{author}{\bibfnamefont{J.}~\bibnamefont{McArthur}}, \bibnamefont{and}
  \bibinfo{author}{\bibfnamefont{R.~F.} \bibnamefont{Siliciano}},
  \bibinfo{journal}{Annu. Rev. Immunol.} \textbf{\bibinfo{volume}{18}},
  \bibinfo{pages}{665} (\bibinfo{year}{2000}).

\end{thebibliography}
\end{document}